\documentclass[10pt,conference,letterpaper]{IEEEtran}
\pdfoutput=1
\usepackage[utf8]{inputenc}
\usepackage{amsmath}
\usepackage{amsfonts}
\usepackage{amssymb}
\usepackage{graphicx}
\usepackage{caption}
\usepackage{subcaption}
\usepackage[protrusion,expansion]{microtype}
\usepackage{url}
\usepackage{xcolor}
\usepackage{verbatim}  
\usepackage[plain,noline,linesnumbered]{algorithm2e}  
\usepackage[colorlinks,linkcolor=blue]{hyperref}
\usepackage{booktabs}
\usepackage{tabularx}
\usepackage{fixltx2e}
\usepackage{listings}
\usepackage{alltt}
\usepackage{tikz}
\usetikzlibrary{automata,arrows.meta,babel,positioning}
\tikzset{shorten >=1pt,node distance=2cm,on grid,
  every state/.style={draw=black,thick,fill=black!20},
  every path/.style={thick},auto,initial text=}

\newcommand{\eg}{\emph{e.g.}}
\newcommand {\eqq}[1]{\textquotedblleft #1\textquotedblright}
\begin{document}

\title{McFSM: Globally Taming Complex Systems}
\author{\IEEEauthorblockN{Florian Murr}
\IEEEauthorblockA{Siemens AG, Corporate Research\\
Otto-Hahn-Ring 6\\
81739 München\\
florian.murr@siemens.com}
\and
\IEEEauthorblockN{Wolfgang Mauerer}
\IEEEauthorblockA{Technical University of Applied Sciences Regensburg \\
Siemens AG, Corporate Research\\
Regensburg and München\\
wolfgang.mauerer@oth-regensburg.de}}

\maketitle


\begin{abstract}
  Industrial computing devices, in particular cyber-physical,
  real-time and safety-critical systems, focus on reacting to external
  events and the need to cooperate with other devices to create a
  functional \emph{system}. They are often implemented with languages
  that focus on a simple, local description of how a component reacts
  to external input data and stimuli.
  Despite the trend in modern software architectures to structure
  systems into largely independent components, the remaining
  interdependencies still create rich behavioural dynamics even for
  small systems.  Standard and industrial programming approaches do
  usually not model or extensively describe the global properties of
  an entire system. Although a large number of approaches to solve
  this dilemma have been suggested, it remains a hard and error-prone
  task to implement systems with complex interdependencies correctly.

  We introduce multiple coupled finite state machines (McFSMs), a
  novel mechanism that allows us to model and manage such
  interdependencies. It is based on a consistent, well-structured and
  simple global description. A sound theoretical foundation is
  provided, and associated tools allow us to generate efficient
  low-level code in various programming languages using model-driven
  techniques.
  We also present a domain specific language to express McFSMs and
  their connections to other systems, to model their dynamic behaviour,
  and to investigate their efficiency and correctness at compile-time.
\end{abstract}

\section{Introduction}
Tackling complex systems usually entails two tasks: Localizing effects to
components of a system, and appropriately treating interdependencies
between the components.
  Most theoretical advances in computer science have been focused
  on the objective of localizing effects, 
  as can be seen in concepts like functional or object-oriented programming
  that try to limit side-effects to manageable portions of the code,
  as do common techniques like \emph{information hiding}
  or \emph{separation of concerns}.
  Most of these are rooted in a solid theoretical basis. 
  The available options for
treating interdependencies in industrial systems are less plentiful.
 
Such systems are often implemented with mechanisms originating
    from programmable logic controllers and finite state machines, or by
    using reactive programming techniques. They
    rely on languages as defined in the ISO EN 61131-3 standard, or employ
    patterns similar to these.
    Finite state machine (FSM)-like approaches, in particular
    sequential function charts (SFC),
    are central to algorithms in these domains.
    FSMs are in widespread use in most areas of computer science:
    Network protocol dispatchers in operating systems, 
    cryptographic handshakes, formal specifications (like UML)
    for embedded systems design~\cite{Borger03}, formalizations of
    representational state transfer (REST) based web applications~\cite{Zuzak11},
    etc.
    All these and many more are based on FSMs.
    The theoretical properties of FSMs are extremely well known.
    Some specialised languages (e.g.,~\cite{Berry92, Halbwachs02}),
    especially reactive languages, are directly based on state machines,
    as are many other approaches
    (e.g. recent languages for intuitive robot control~\cite{Diewald16}).
    A considerable number of extensions to FSMs,
    like hierarchical state machines~\cite{OMG09},
    some based on the seminal statechart idea~\cite{Harel87}, have been suggested.

Albeit libraries to implement FSMs and related approaches are 
    available for most languages, most of them do not provide
    explicit expressive constructs to describe the
    global structure of composite systems.
    \footnote{We deliberately ignore
  regular expressions. These are in fact an integral part of many
  languages and stem from FSM based origins, but are used as acceptors
  of languages without making the underlying state machines accessible.} 
    Using the state machine pattern~\cite{Gamma95}, FSMs
    are usually
    hand-crafted with generic language techniques,
    for instance using \texttt{case} statement dispatchers in C
    or similar constructs.
    This way, it is easy to (inadvertently) mix the pure simplicity
    of the FSM-approach with Turing-complete language features, thus
    sacrificing compile-time provability.
    The same observation holds for Turing-complete reactive programming systems.

For distributed systems, the description of
    the state machine
    that comprises the system is often spread across multiple physical
    or virtual machines or at least multiple \emph{local} program components, 
    making it hard to obtain a consistent and coherent picture of the 
    \emph{global} system that is required to guarantee system-wide 
    properties, like overall safe behaviour.
    The level of diversity in industrial cyber-physical systems is still
    substantially more pronounced than the diversity in, say,
    operating systems, tools, programming languages etc.
    This suggests that
    despite the previously suggested multitude of approaches,
    the practical, real-world aspects
    of the field are far from being solved satisfactorily.

This paper introduces a novel mechanism -- multiple coupled finite
    state machines (McFSMs) -- together with a domain specific
    language (DSL) for model-based development and abstract specification.
The mechanism uses multiple FSMs coupled by notifications to 
    make the structure of a cooperating system explicit. It aims at 
    retaining the simplicity and advantages of local FSMs by enveloping 
    them with a global superordinate structure that takes care of the 
    intricacies brought about by their interdependencies. 
As a \emph{low-level} mechanism, it lends itself to an efficient 
    implementation. As a \emph{global superordinate} structure, it avoids 
    the pitfalls of scattered observers acting largely agnostic of their 
    interdependencies. Owing to a \emph{theoretical} foundation, it lends 
    itself well to rigorous scrutiny and formal verification. 

McFSMs aim primarily at explicitly describing and handling
    interdependencies between components at compile-time.  They also
    have a thorough theoretical basis that allows us to prove various
    system properties. McFSMs can be used to reduce implicit or undesired dependencies
    between components and foster a consistent and
    well-structured global description of interdependencies.
    The approach can also serve as
    basis for deterministic hard real-time systems, and is therefore
    applicable to a wide class of industrial systems.

Finite state machines are one of the earliest theoretical concepts in
    automated computing, and also form the basis of many modern software
    engineering mechanisms like UML state machines~\cite{OMG09}.
    Code
    for (real-time) systems can be synthesized from FSM based
    descriptions (see, e.g.,~\cite{Dietrich15,Halbwachs02}).
    It is
    known that when systems comprising multiple components are
    modelled using a straightforward product automaton approach, an
    exponential increase in both the number of states and edges can
    occur, which makes the approach inherently unfeasible for
    practical software engineering purposes.
    It creates very dense
    and confusing diagrams even for systems of moderate complexity,
    or might use up to much space.
    Our
    approach
    allows
    to describe industrially relevant
    coupled systems while avoiding such an exponential
    \eqq{state space explosion}.

\section{Modelling Interacting Systems}
Systems based on interacting components, either logical or physical,
    need to propagate information about state changes between their
    constituents. When a state change occurs in one part of the
    system, the change propagates to related components, and may trigger
    further state changes.  This notion is generically captured by the
    \emph{observer pattern}~\cite{Gamma95}: An object \(A\)
    (the \emph{subject}) maintains a list of dependent objects (the
    \emph{observers}) \(\{B_{1}, B_{2}, \ldots, B_{n}\}\)
    and notifies them when any state change takes place in object \(A\).
     
    There is growing evidence that the observer pattern is problematic
    (see, for instance, Ref.~\cite{Maier12}).  Industrial experience
    endorses these findings with the observation that modelling even
    superficially extremely simple systems, for instance displays for
    multi-function household appliances with a small number of
    controls, often results in complex and error-prone systems that
    require substantial implementation and testing efforts when the
    architecture is based on the observer pattern. The authors are
    aware of industrial projects where the initial estimated effort of a
    few weeks resulted in several months worth of implementation
    effort.

We want to emphasise four major challenges: Firstly, observers promote
    side-effects since their states are only implicitly available, but not
    explicitly represented by programming language constructs.  Shared
    states need therefore be made available in a context that is
    accessible from multiple observers, or even globally,
    of course
    without violating information encapsulation principles. 
    Secondly,
    observers can execute arbitrary Turing-complete code, which can easily
    lead to violating the principle of separation of concerns. 
    Thirdly,
    traditional programming languages make it hard or even impossible to
    statically analyse and understand the dynamic control flow when chains
    of dynamically registered observers are used.
    Finally, any state
    change apart from the one in subject \(A\) triggering the notifications is
    not part of the observer pattern and is therefore ``invisible'' from
    its point of view.

Nonetheless, the observer pattern enjoys wide-spread use in many
    software systems, and likewise do the very similar publish-subscribe and
    signal-slot mechanisms. Handling asynchronous interrupts on the system
    level is also closely related to the described mechanisms.

Possible solutions to the aforementioned problems include the use of
    reactive programming techniques~\cite{Maier12} that require only a
    specification of dependencies between interacting components, as
    compared to manual encoding execution flows. However, such
    techniques often require intensive run-time support and advanced language
    features that are not available in languages conventionally used for
    systems programming.

\section{Multiple Coupled FSMs}
\subsection{Example and DSL}
To facilitate the practical use of McFSMs, we provide a 
    domain-specific language (DSL).
    An example to illustrate the language is based on an
    application pattern that occurs frequently in automation and control.
    Consider a distributed set of binary (on/off) switches that reside
    at different locations in a plant or a residential building;
    triggering one of the switches changes the state of
    one shared dependent entity, perhaps a light bulb (on/off) or a
    status indicator (green/yellow/red).
    Even for a system as simple as two binary switches and
    one ternary status indicator, a straightforward FSM approach
    leads to \(2^{2}=4\) possible combinations of switch settings
    (off/off, on/off, off/on, and on/on) and
    three different values for the indicator resulting in
    \(3\times 4=12\) states.
    For each state, two edges describing the result of
    flipping the switch are necessary.
    For the more general case of \(n\) switches and an indicator
    with \(m\) levels, the amount of states is \(m\times 2^{n}\).  
    The FSM grows exponentially in the number of switches, 
    which obviously makes it hard to obtain an intuitive
 understanding
    of the problem from whatever graphical representation 
    might be chosen. 

\begin{figure}[htb]
      \begin{center}\begin{tikzpicture}
          \node[state,initial] (s0_0)                 { 0 };
          \node[state]         (s0_1) [right of=s0_0] { 1 };
          \node[state,initial] (s1_0) [right of=s0_1] { 0 };
          \node[state]         (s1_1) [right of=s1_0] { 1 };
          \node[state,initial] (l0_0) [below of=s0_0] { G };
          \node[state]         (l0_1) [right of=l0_0] { Y };
          \node[state]         (l0_2) [right of=l0_1] { R };

          \path[->] (s0_0) edge [bend left] node [below] { \(f_{0}\) } (s0_1) 
                    (s0_1) edge [bend left] node [below] { \(f_{0}\) } (s0_0) 
                    (s1_0) edge [bend left] node [below] { \(f_{1}\) } (s1_1) 
                    (s1_1) edge [bend left] node [below] { \(f_{1}\) } (s1_0) 
                    (l0_0) edge [bend left] node [above] { \(f_{0}, f_{1}\) } (l0_1) 
                    (l0_1) edge [bend left] node [above] { \(f_{0}, f_{1}\) } (l0_2) 
                    (l0_2) edge [bend left=45] node [above] { \(f_{0}, f_{1}\) } (l0_0) 
                    ;
        \end{tikzpicture}\end{center}
      \caption{McFSM visualization of a ternary status indicator
        controlled by two binary switches. The exponential state space
        explosion of traditional FSMs is replaced by a linear growth
        in the number of states (and edges) depending on the (increasing) number
        of switches.}\label{fig:mcfsm}
    \end{figure}
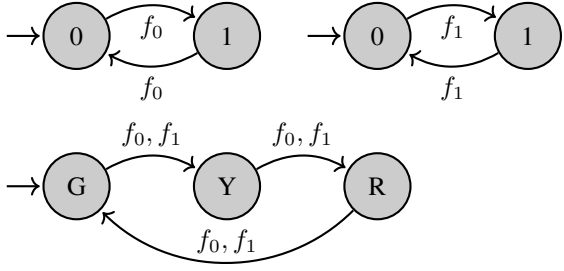

Now in contrast how our formalism describes the 
    scenario: Each binary switch \(S_{i}\) 
    is modelled with two states, and the ternary indicator \(I\) 
    with three states, as shown in Figure~\ref{fig:mcfsm}. All 
    couplings between the switches and the indicator are introduced by 
    edge labels: Labels \(f_{i}\) 
    \emph{below} the edges of switch \(i\) 
    emit event \(f_{i}\) 
    (a flip of switch \(i\)) 
    that is consumed by the ternary indicator, as indicated by the 
    labels \emph{above} the edges.
    Thus \(I\) cyclically switches its state 
    in response to these events. The visual representation of the 
    scenario is as shown in Figure~\ref{fig:mcfsm}.
    
\begin{small}
\lstset{language=tcl} 
\begin{lstlisting}
FSM class "HealthSignal" {
    hop green_yellow  += xFlip yYellow
    hop yellow_red    += xFlip yRed
    hop red_green     += xFlip yGreen
}

FSM class "Switch" {
    hop up_down  += xPress yFlip
    hop down_up  += xPress yFlip
}

McFSM class "ComboSwitches" {
    Switch inst S1 {
        Start: up
        cap &xPress  += ../xPressS1
    }
    Switch inst S2 {
        Start: up
        cap &xPress  += ../xPressS2
    }
    HealthSignal inst Lights {
        Start: yellow
        cap &xFlip   +=  ../S*/yFlip
    }
}
\end{lstlisting}
\end{small}

The DSL
    %
    distinguishes between classes and instances,
    which is familiar to most programmers.
    Classes are very similar to traditional FSMs (as indicated
    by the keyword \texttt{FSM}) in that they are characterised
    by their states and transitions.
    However, they do not require an initial state.
    We distinguish between \emph{incoming} transitions
    (identifiers prefixed by ``\texttt{x}'') 
    triggered by external events, and \emph{outgoing} transitions 
    (identifiers prefixed by ``\texttt{y}'').\footnote{The 
    rationale for using prefixes \texttt{x} and \texttt{y} is the form 
  \(y=f(x)\) 
  of algebraic equations, where \(x\) 
  is an input and \(y\) 
  an output value. In the visual representation, this corresponds to 
  labels above and below the edges.}

\emph{Edges} correspond to state transitions (\texttt{hop}s)
constructed e.g. by \texttt{hop up\_down += xPress yFlip}: This
generates a directed edge from state ``up'' to state ``down'', and
labels it with \texttt{xPress} and \texttt{yFlip}.  Thus this
transition is triggered in state up when event xFlip is observed (when
the button is pressed), causing the edge-id to be added to an internal
list of events that will be processed.  The output \texttt{yFlip} may
be used in the DSL to group or abstract from concrete edge-ids.  Note
that edge constructions implicitly create states as well.

    Classes, instances, states, edges and events can be hierarchically
    nested. If an object \texttt{obj2} is hierarchically below
    \texttt{obj1}, the DLS represents this using the syntax
    \texttt{/obj1/obj2}. We cannot fully describe the concept here,
    but in the above example, the McFSM class \texttt{ComboSwitches}
    is on level 1 of the global hierarchy, and so are the input events
    \texttt{xPressS1} and \texttt{xPressS2}.  FSMs instantiated as
    part of the McFSM reside in level 2, and so on. The syntactic
    element \texttt{../} refers to the previous hierarchy level.
    
    Connecting edges in FSM instances with events requires a generic
    syntax that can select instances of interest. Consider, for
    instance, the statement \texttt{cap \&xPress += ../xPressS1}: It
    selects every instance of \texttt{/ComboSwitch/S1} that already
    has an event annotation \texttt{xPress}, and adds an additional
    annotation \texttt{../xPressS1} (in absolute form:
    \texttt{/ComboSwitch/xPressS1}). This connects the input event
    \texttt{xPressS1} with the relevant transitions in the button
    instance. Every time a (possibly physical) button,
    mapped to the external McFSM event by system mechanisms, is
    pressed, the appropriate transitions are triggered in the class
    instance that models the switch.

In general, \texttt{cap} expects an edge list as parameter, and adds 
labels to these edges using the \texttt{+=} operator. Edge lists and 
lists of labels may be specified in three ways: 1) As a list of 
absolute/relative edges/labels, 2) using glob patterns as in 
\texttt{../S*/yFlip} (this selects all \texttt{yFlip} labels on switch 
instances \texttt{S1} and \texttt{S2}), and 3) as semantic references 
of the form \texttt{\&xFlip}, which selects the set of all edges labelled 
with \texttt{xFlip}.



\subsection{Formalism}
\newcommand{\exts}{\text{ext}}
\newcommand{\ints}{\text{int}}
An FSM \(F\)
    is given by a four-tuple \(F=(Q, \Sigma, \delta, q^{(0)})\),
    where \(Q\)
    is a finite set of states, \(\Sigma\)
    is a finite set of events, \(\delta: Q\times \Sigma \rightarrow Q\)
    denotes the transition function that determines the next state given
    the current state and an input event, and \(q^{(0)}\in Q\)
    is the initial state of the machine.

A McFSM \(M = (Q, \Sigma, \delta, q^{(0)})\)
    comprises a set of $n>1$ interacting FSMs
    $\{F_{1}, F_{2}, \ldots, F_{n}\}$,
    with state space $Q = Q_{1} \times \cdots \times Q_{n}$,
    event set $\Sigma = \Sigma_{\exts} \cup \Sigma_{\ints}$
    and initial state $q^{(0)} = (q_1^{(0)}, \ldots, q_n^{(0)})$.

The events in $\Sigma_{\exts}$ constitute the ``interface'' of the
McFSM and can be connected, for instance, to inputs of physical sensors
as parts of a larger program. $\Sigma_{\ints}$ is invisible to the outside.
It consists
    of pairs of states $(q_i^j,q_i^k) \in Q_i$ that
    create \emph{couplings} between the FSMs $F_i$,
    because the \emph{internal event} $e_{i,j,k} = (q_i^j,q_i^k)$
    occurs every time the state transition 
    $q_i^j \mapsto q_i^k$ occurs in $F_i$, 
    but can be associated with any $F_x$
    by defining $\delta(q_x,e_{i,j,k})$ ($q_x \in Q_x$),
    thus notifying the observing FSM $F_x$.

Operationally, events $a \in \Sigma_{\exts}$
    can be
    provided by arbitrary system sources;
    their processing by a McFSM
    generates internal coupling events $b \in \Sigma_{\ints}$,
    because \emph{state
    transitions themselves are considered to be events}.

The transition function $\delta$ of the McFSM $M$ combines the transition
    functions $\delta_{1}, \ldots, \delta_{n}$ of the constituent FSMs by
    providing any input event $a$ 
    and the ensuing coupling events $b$
    to the machines \(F_{i}\)
    in a predefined order that may be event-dependent,
    but usually just follows the order of the machines $F_1, \ldots, F_n$.
   
    To realise the described event distribution and coupling, we use a
    data structure called \emph{XQueue} that combines queue- and
    stack-like features and enforces that 1) all events are treated in
    the order they occur, as in a queue, and 2) all ensuing
    \emph{coupling events} are treated before the next event \(a\)
    is processed, as in a stack. Processing is performed as an 
    atomic step before any side effect handlers are called. A precise
    formulation of the algorithm is given in the accompanying website
    and the system source code.\footnote{To be published under an open
      source license.}

The UI can show an upper bound to the amount of steps
    required to distribute an external event to the components, and
    execute their actions. This makes the formalism
    suitable for real-time systems.


\section{Relation to other approaches}
McFSMs share many desirable properties with ordinary FSMs, in 
    particular the ability to prove predicates on their state-spaces,
    which is relevant for safety-critical systems.  

    In the observer pattern, the observers 
    \(B_{1}, B_{2}, \ldots, B_{n}\) 
    can again be subjects of further observers.  All these 
    objects can be viewed as multiple FSMs that are 
    \emph{coupled} through \emph{notifications}, that is, function 
    calls of \emph{event-handlers}. These cascading notifications can 
    induce substantial hidden complexity that can only be tested at 
    \emph{run-time}. A McFSM combines the state spaces of the 
    constituent FSMs, but makes coupling events (observer 
    \emph{notifications}) explicit at \emph{compile-time}, which 
    provides the basis for debugging and proving static correctness 
    properties. A McFSM handles state transitions as one single 
    \emph{atomic} action and then calls event handlers that can 
    perform any 
    necessary side-effects. 

    While the available space does not permit to discuss all 
    differences and similarities with previous approaches in detail,
    we note that McFSMs share the idea of dividing a system into 
    sub-automata with UML hierarchical state machines (our ability to 
    generate instances from classes provides a template-like extension 
    to the idea). Likewise, the concept of emitting signals from edge 
    transitions also appears in statecharts and related formalisms. 

    As for the differences, we would like to highlight two salient 
    points: Firstly, our pattern matching based interdependency 
    specification does not only lead to considerably increased 
    expressive power, but further economises the number of arrows 
    appearing in charts -- our formalism can be seen as hybrid between 
    a reactive programming language (intentionally not 
    Turing-complete) and a visual modelling mechanism where both 
    aspects complement each other. Secondly, McFSMs (in particular 
    using the XQueue mechanism) have been designed from the ground up 
    to provide well-defined mathematical semantics, which eliminates 
    one major criticism (see, \eg,~\cite{Derler12}) of statechart-like 
    approaches. 

    Another related approach, the Labelled Transition System Analyser
    (LTSA, see~\cite{Magee00}), aims specifically at modelling
    concurrency, and proving certain properties of concurrent systems.
    Although LTSA is also centered around appropriately combining
    finite state machines, our approach differs in that concurrency is
    not the core focus, which is also reflected in the substantially
    different description language that aims on system design, not at
    describing concurrent processes.  The language also avoids mixing
    any side-effects into the description language, and
    strives
    that
    expressive power is combined with utmost syntactic and conceptual
    simplicity, which is especially desirable in efforts that require
    (safety or other types of) certification.
    Despite the
    impossibility of making such statements objective, we have tried
    to leverage decades of industrial experience to achieve the goal
    to the best of our abilities.
   

\section{Tools \& Outlook}
The approach described in this paper is accompanied by an integrated 
set of tools and a graphical user interface (GUI) that combines the DSL with 
a visual representation, and can be used to experiment with our new
approach
. More information is available on the 
supplementary website 
\url{https://hps.hs-regensburg.de/maw39987/icse/icse.html}.

Future work will focus on testing the efficiency of the specification
language and the GUI on practical problems from various domains:
Using experts
from the fields, we intend to collect typical problems, and implement
solutions using the McFSM formalism.
By carefully evaluating the results, we will further improve the DSL,
the GUI and other tools involved, especially regarding their intuitive
usability.
    


\end{document}